# Spatiotemporal mode-locked vector solitons

Jia-Wen Wu[1], Rong-Jun Huang[1], Jia-Hao Chen[1], Hu Cui[1], Zhi-Chao Luo[1], Wen-Cheng Xu[1], Xiao-Sheng Xiao[2,*], Ai-Ping Luo[1,*]

[1] Guangdong Provincial Key Laboratory of Nanophotonic Functional Materials and Devices, South China Normal University, Guangzhou, Guangdong 510006, China

[2] State Key Laboratory of Information Photonics and Optical Communications, School of Electronic Engineering, Beijing University of Posts and Telecommunications, Beijing 100876, China

*Corresponding authors: xsxiao@bupt.edu.cn, luoaiping@scnu.edu.cn

**Abstract**: With the increased transverse mode degrees of freedom, spatiotemporal mode-locked (STML) fiber lasers exhibit more intricate and richer nonlinear dynamics, making them an ideal platform for studying complex nonlinear phenomena. However, current research mainly focuses on their scalar characteristics, leaving their vector characteristics unexplored. Here, we investigate the vector characteristics of the STML fiber laser and demonstrate two novel types of vector solitons associated with transverse modes, namely the STML polarization-locked vector soliton (PLVS) and the STML group velocity-locked vector soliton (GVLVS). In both types of STML vector solitons, the two polarization modes exhibit distinct transverse mode compositions and relative power ratios. However, the two polarization modes share identical peak wavelengths in STML PLVSs, while they have different peak wavelengths in STML GVLVSs. Notably, during the soliton splitting process of the STML GVLVSs, polarization-dependent phenomena, including the gain competition and variation of the peak wavelength difference between polarization modes as well as the invisible periodic variation in the beam profile, are observed. The formation of STML vector solitons demonstrates that soliton trapping remains a universal phenomenon for vector solitons even in the more intricate STML fiber lasers, and the obtained results reveal the vector characteristics of STML fiber lasers, enhancing the understanding of their nonlinear phenomena.

## 1. Introduction

As a type of high-dimensional mode-locked fiber laser, the spatiotemporal mode-locked (STML) fiber laser based on multimode fibers, where multiple transverse modes and longitudinal modes are simultaneously locked, has attracted widespread interest. Due to the large core diameter of multimode fibers, they can support higher pulse energy and more transverse modes, making STML fiber lasers a promising option for generating high-energy pulses and investigating complex high-dimensional nonlinear phenomena. In 2017, Wright et al. demonstrated the first STML fiber laser using multimode fibers, and the high-energy and high-dimensional STML solitons were achieved [1]. In order to establish spatiotemporal mode-locking, the spatial filtering effect was deliberately introduced in the laser to counteract modal dispersion. Since the compensation of modal dispersion plays a critical role in establishing spatiotemporal mode-locking, the STML fiber lasers with larger modal dispersion were constructed and investigated to further explore the underlying mechanisms. Remarkably, even in such lasers with large modal dispersion, the spatial filtering effect still plays a pivotal role in counteracting mode dispersion to achieve spatiotemporal mode-locking [2]. The weak spatial filtering effect is conducive to obtaining multimode Q-switched pulses, while the strong spatial filtering effect favors the achievement of STML pulses [3]. Furthermore, the spatial filtering effect is also the key effect in counteracting the modal dispersion in the mid-infrared STML fiber laser [4]. The multimode interference filtering effect is a type of filtering effect induced by differences in fiber core diameters. Since it can provide both spectral and spatial filtering effects, it was widely utilized in STML fiber lasers to counteract modal dispersion and chromatic dispersion [5, 6]. In addition to the spatial filtering effect, the saturable absorption effect is also identified as the key effect in compensating the modal dispersion to achieve spatiotemporal mode-locking [7]. In summary, the formation of STML dissipative solitons is based on the delicate balance among chromatic dispersion, modal dispersion, nonlinearity, spectral filtering effect, spatial filtering effect, saturable absorption effect, gain, and loss. By regulating these intracavity effects, various STML solitons, including multiple solitons [8], soliton molecules [9], self-similar solitons [10], dispersion-managed solitons [11], dissipative soliton resonances [12], and noise-like solitons [13], were achieved. Due to the complex balance among these intracavity effects and interactions among transverse modes, the STML solitons exhibit intricate and rich characteristics. Different transverse modes were observed to possess distinct spectra and peak wavelengths [8, 9]. Furthermore, the dynamics of STML multiple solitons [14, 15], soliton molecules [15, 16], and pulsating solitons [15, 17, 18] were found to be more intricate and mode-dependent.

On the other hand, due to the imperfect circular symmetry of the fiber cores as well as stress and bending on the fibers, fibers always exhibit birefringence and thus actually support two orthogonal polarization modes. However, due to the birefringence, these two polarization modes have different group velocities and phase velocities, resulting in polarization modal dispersion. When these two polarization modes are also trapped together and propagate at the same group velocity, a vector soliton is formed. Therefore, compared to the scalar soliton that neglects birefringence, the formation of vector solitons in single-mode fiber lasers necessitates balancing not only chromatic dispersion, nonlinearity, gain, and loss but also polarization modal dispersion. By regulating these effects, diverse types of vector solitons, including polarization-locked vector solitons (PLVSs) [19], group velocity-locked vector solitons (GVLVSs) [20], and polarization-rotating vector solitons (PRVSs) [21], could be achieved. Since vector solitons contain two polarization modes, they exhibit more complex and richer characteristics than scalar solitons. In PLVSs, the two polarization modes share identical peak wavelengths [22, 23], while in GVLVs, they exhibit distinct peak wavelengths [24, 25]. Different from them, the two polarization modes in PRVSs not only possess distinct peak wavelengths but also exhibit periodically varying intensities [22, 23]. Furthermore, different polarization modes within vector solitons were also observed to manifest distinct dynamics, such as the soliton pulsating dynamics [26]. In contrast to single-mode fiber lasers, to form STML vector solitons in multimode fiber lasers, not only multiple longitudinal modes and polarization modes but also multiple transverse modes are required to be simultaneously locked. Thus, the formation of STML vector solitons requires a delicate balance among chromatic dispersion, nonlinearity, gain, loss, polarization modal dispersion, modal dispersion, spectral filtering effect, spatial filtering effect, and saturable absorption effect. Since STML vector solitons involve multiple transverse modes and polarization modes, they will exhibit rich and fascinating characteristics despite their complex formation. However, current research on STML fiber lasers primarily concentrates on the characteristics of STML scalar solitons, and studies on STML vector solitons have not yet been reported.

In this work, we investigate the vector characteristics of the STML fiber laser and demonstrate two novel types of vector solitons associated with transverse modes, namely the STML PLVS and the STML GVLVS. In both types of STML vector solitons, polarization-resolved measurements confirm distinct transverse mode compositions and relative power ratios for two orthogonal polarization modes. However, the peak wavelengths of the two polarization modes in STML PLVSs are the same, while those in STML GVLVSs are different. By adjusting the pump power and polarization controller (PCs), the single soliton, multiple solitons, and harmonic solitons are achieved in both STML PLVS and STML GVLVS. During the soliton splitting process of the STML GVLVSs, gain competition between polarization modes and the invisible periodic variation in the beam profile are observed. In addition, the peak wavelength difference between the two polarization modes decreases as the number of solitons increases. Benefiting from the combined filtering effects in the cavity, both wavelength-tunable STML PLVSs and STML GVLVSs are achieved. The formation of these two types of STML vector solitons demonstrates that soliton trapping remains a universal phenomenon for vector solitons even in the more complex STML fiber lasers, and the obtained results reveal their vector characteristics, contributing to a deep understanding of the STML fiber lasers.

## 2. Experimental setup

The schematic of the proposed STML vector soliton fiber laser is illustrated in Fig. 1. A 2.08 m long erbium-doped fiber (EDF, Er1200-20/125, FORC, RAS) is employed as the gain medium and is pumped by a 980 nm laser diode via a wavelength division multiplexer (WDM). The fiber pigtails (FUD-3729, MM-GSF-20/125-10A) of the WDM have the same core diameter of 20 μm as the EDF, and both support six transverse modes. Since polarization discrimination components will fix the polarization of the light and prevent the formation of vector solitons, a polarization-independent semiconductor saturable absorber mirror (SESAM) is served as the saturable absorber to achieve vector solitons. In addition, this SESAM also functions as a reflector to form a linear cavity with a Sagnac loop at the opposite end of the laser. The Sagnac loop consists of a 2×2 optical coupler (OC) with a splitting ratio of 20:80 and a PC1. Beyond functioning as a reflector, the Sagnac loop also exhibits filtering effect, so the PC1 in it can be employed to adjust the filtering characteristics of the laser. Except for the EDF and the pigtails of the WDM, all other fibers within the cavity are the six-mode fibers (YOFC, FM2011-A) with a core diameter of 23 μm, and the total cavity length is approximately 6.64 m. Due to the differences in fiber core diameters, there exists the multimode interference filtering effect in the cavity [27, 28]. Therefore, PC2 can also be used to adjust the filtering characteristics of the laser. In addition to adjusting the filtering characteristics, both PC1 and PC2 are also employed to adjust the polarization of the light and the intracavity birefringence, thereby tuning the characteristics of the STML vector solitons. The laser is extracted via the 2×2 OC. In order to further investigate the vector characteristics of the solitons, another PC3 and a six-mode-fiber-based polarization beam splitter (PBS) are externally connected to the OC for polarization-resolved measurements. The spatiotemporal and vector characteristics of the output solitons are measured by an optical spectrum analyzer (Yokogawa, AQ6317C) with 0.01 nm resolution, a 13 GHz oscilloscope (Teledyne Lecroy, 813Zi-B), a radio-frequency (RF) spectrum analyzer (Agilent, E4407B ESA-E SERIES), an autocorrelator (Femtochrome, FR-103WS), and a charge coupled device camera (Goldeye G-033SWIRTEC1).

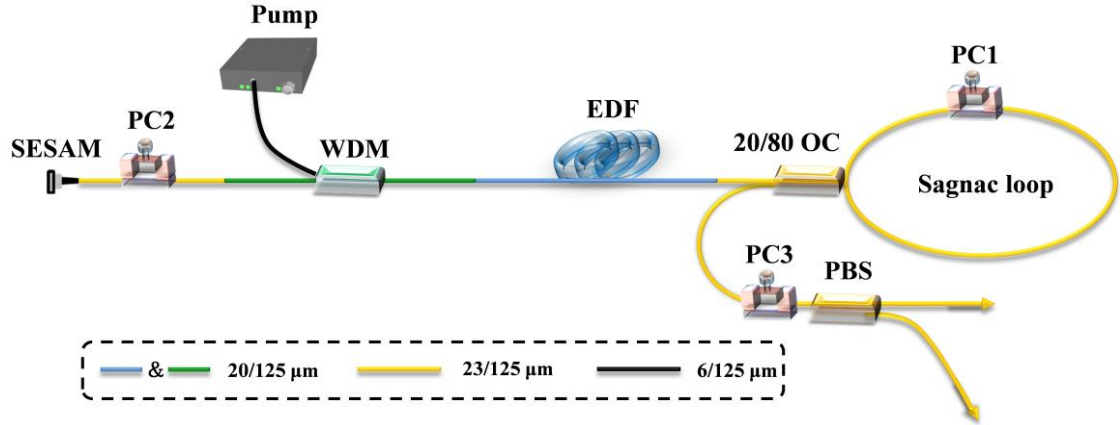

**Figure 1.** Schematic of the STML vector soliton fiber laser. Different types of fibers are color-coded for identification.

## 3. Experimental results
### 3.1 STML PLVS operation

When the PC1 and PC2 are appropriately set, self-starting STML vector solitons can be easily achieved by increasing the pump power to approximately 130 mW. Figure 2 presents the characteristics of the STML vector single soliton at a pump power of 159 mW. We first measure the characteristics of this soliton directly output from the laser through the 2×2 OC and before passing through the PBS. This STML vector soliton exhibits a central wavelength of 1560.29 nm and a 3 dB spectral bandwidth of 0.10 nm, as indicated by the red curve in Fig. 2(a). The corresponding pulse-train is shown by the red line in Fig. 2(b). The interval between the adjacent pulses measures 64.2 ns, which is consistent with the round-trip time for the light to travel in the cavity, indicating that the laser is operating in the single-soliton state at the fundamental repetition rate. As displayed in Fig. 2(c), a peak with a signal-to-noise ratio of 58.0 dB is located at 15.58 MHz in the RF spectrum with a span of 500 kHz, further confirming the operation at the fundamental repetition rate. Additionally, we also measured the RF spectrum with a range of 1 GHz, as shown in the inset of Fig. 2(c). No significant intensity modulation is observed, indicating that the laser operates stably. The measured autocorrelation trace given in Fig. 2(d) exhibits a full width at half maximum of 81.0 ps. Assuming a Sech$^2$ shape, the pulse width is calculated to be 52.6 ps. The resulting time-bandwidth product is 0.648, which suggests that this STML vector soliton has a certain chirp. The measured beam profile shown in Fig. 2(e) exhibits a spatial intensity distribution significantly deviating from that of the fundamental mode, indicating the presence of high-order transverse modes within the soliton.

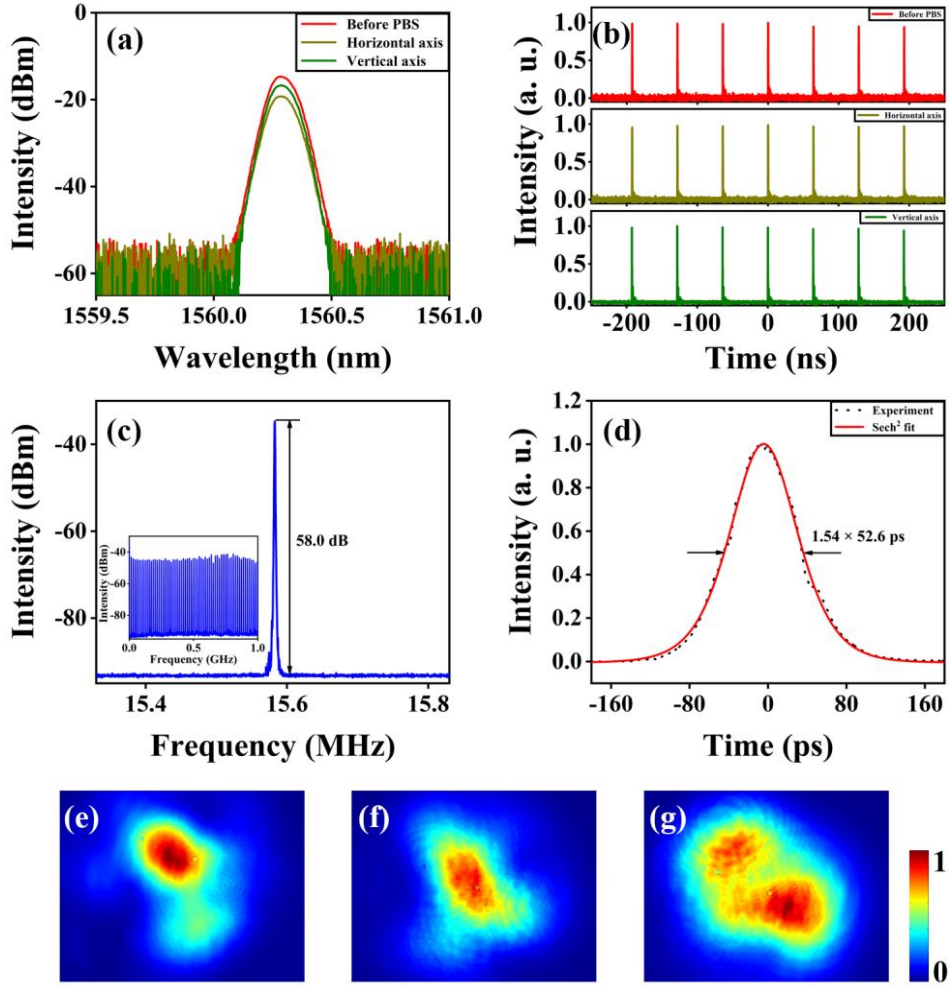

**Figure 2.** Characteristics of the STML polarization-locked vector single soliton. (a) The optical spectra before (red line) and after (yellow and green lines) passing through extra-cavity PBS; (b) pulse-trains before (red line) and after (yellow and green lines) passing through extra-cavity PBS; (c) RF spectrum with a span of 500 kHz before PBS (inset: RF spectrum with a span of 1 GHz before PBS); (d) autocorrelation trace before PBS; (e) beam profile before PBS; (f) beam profile of the polarization mode on the horizontal axis; and (g) beam profile of the polarization mode on the vertical axis.

To investigate the vector characteristics of the soliton, an extra-cavity PC and a PBS are connected to the output of the laser, and its polarization-resolved characteristics are measured at both output ports of the PBS. The yellow and green curves in Fig. 2(a) display the optical spectra of two orthogonal polarization modes. These two polarization modes exhibit distinct intensities, with the output power of the polarization mode on the horizontal axis measured to be 321 μW and that of the polarization mode on the vertical axis measured to be 475 μW. However, their spectral profiles are similar and closely match that before PBS. Moreover, they also have the same peak wavelength of 1560.29 nm and the same 3 dB spectral bandwidth of 0.10 nm, consistent with those before PBS. Their corresponding pulse-trains, indicated by the yellow and green lines in Fig. 2(b), both exhibit uniform intensity. Furthermore, we also measured the beam profiles of two polarization modes and give them in Fig. 2(f) and (g). Notably, the distinct spatial intensity distributions of the two polarization modes indicate that they have different transverse mode compositions and relative power ratios, resulting in the modal dispersion between them. Thus, the formation of this STML vector soliton also involves the compensation of modal dispersion between different polarization modes. Based on the formation mechanisms of vector solitons in single-mode fiber lasers and STML scalar solitons in multimode fiber lasers, the formation mechanism of STML vector solitons can be summarized as a delicate balance among chromatic dispersion, modal dispersion, polarization modal dispersion, nonlinearity, spectral filtering effect, spatial filtering effect, saturable absorption effect, gain, and loss. Therefore, the identical peak wavelengths for different polarization modes within the STML vector soliton in Fig. 2 can be attributed to the fact that a delicate balance among these effects is achieved. This equilibrium eliminates the need to introduce additional chromatic dispersion between the two polarization modes by shifting their peak wavelengths to form the vector soliton. This STML vector soliton demonstrates identical characteristics in the optical spectra and temporal pulses to the PLVS in single-mode fiber lasers. However, its two polarization modes contain distinct transverse mode compositions and relative power ratios, involving locking of multiple transverse modes. Therefore, we refer to this type of STML vector soliton as the STML PLVS.

When the nonlinear phase exceeds the critical threshold, the soliton will split, resulting in the formation of multiple solitons. In our laser, STML polarization-locked vector multiple solitons can be generated by adjusting the pump power and PCs. Figure 3 presents the characteristics of the STML polarization-locked vector dual solitons. The optical spectrum before passing through the PBS exhibits a central wavelength of 1560.41 nm, as indicated by the red curve in Fig. 3(a). The corresponding pulse-train is shown by the red line in Fig. 3(b). There exist two pulses with a pulse interval of 10.9 ns within a round-trip time of the cavity. The beam profile in the inset of the upper panel of Fig. 3(b) demonstrates the presence of high-order transverse modes. The spectra of the horizontal and vertical polarization modes after passing through the PBS are displayed by the yellow and green curves in Fig. 3(a), respectively. They have the same peak wavelength as that before passing through the PBS despite the intensity differences. Their pulse-trains are represented by the yellow and green lines in Fig. 3(b), respectively. As observed before PBS, both have two pulses with a pulse interval of 10.9 ns within a round-trip time of the cavity, and the pulse intensity is uniform for each round trip. The corresponding beam profiles are presented in the insets of the middle and bottom panels of Fig. 3(b). Although both polarization modes display dual-lobe-shaped beam profiles, their spatial intensity distributions exhibit difference, indicating the existence of modal dispersion between the two polarization modes. Consequently, the formation of the STML polarization-locked vector dual solitons also involves the compensation of modal dispersion between different polarization modes. In addition, STML polarization-locked vector triple solitons and quadruple solitons are also achieved.

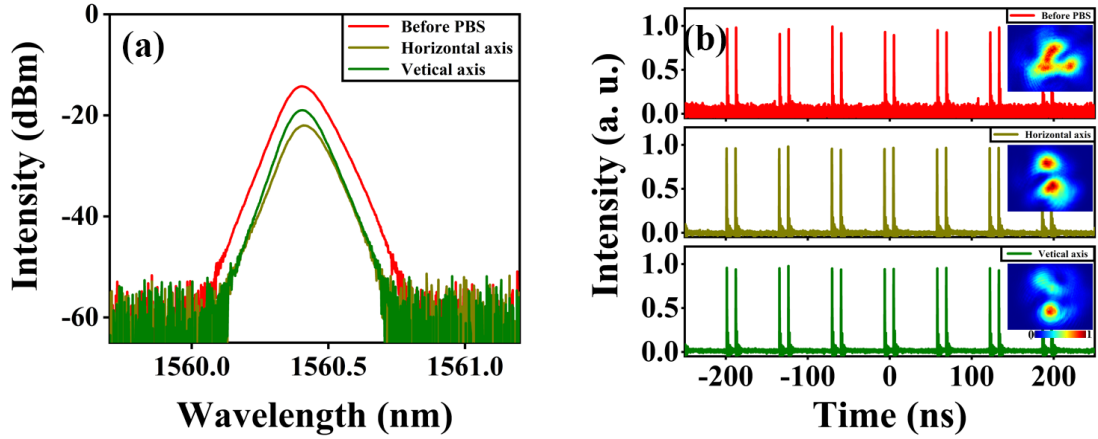

**Figure 3.** Characteristics of the STML polarization-locked vector dual solitons. (a) The optical spectra before (red line) and after (yellow and green lines) passing through PBS; and (b) pulse-trains before (red line) and after (yellow and green lines) passing through PBS (insets: corresponding beam profiles).

As a special type of multiple solitons, harmonic solitons exhibit equally spaced pulse-trains, with the pulse spacing being equal to integer fractions of the round-trip time of the cavity. By carefully adjusting the pump power and PCs, STML polarization-locked vector harmonic solitons can be achieved in our laser. Figure 4 presents the characteristics of the STML polarization-locked vector second harmonic solitons. The optical spectra before and after passing through the PBS share an identical peak wavelength of 1560.29 nm, as shown in Fig. 4(a). Their pulse trains are given in Fig. 4(b), and they have a uniform intensity. In addition, they all have the same pulse spacing of 32.1 ns, which is half of the round-trip time of the cavity, confirming second harmonic solitons operation. The corresponding beam profiles presented in the insets of Fig. 4(b) exhibit distinct spatial intensity distributions, indicating that the formation of the STML polarization-locked vector harmonic solitons also involves the compensation of modal dispersion between different polarization modes. Figure 4(c) presents the measured RF spectrum. There is almost no peak at the fundamental repetition rate of 15.58 MHz, but a peak with a signal-to-noise ratio of 45.3 dB appears at twice the fundamental repetition rate (31.16 MHz). This further confirms that the laser operates in the state of second harmonic solitons. In addition to the STML polarization-locked vector second harmonic solitons, third harmonic solitons and fourth harmonic solitons are also obtained in our laser.

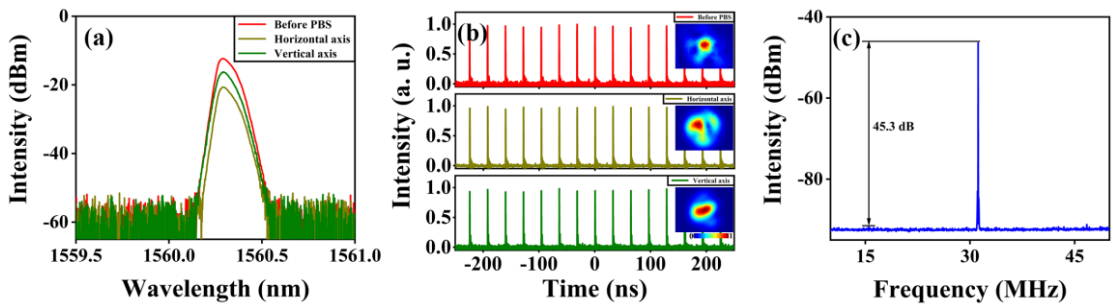

**Figure 4.** Characteristics of the STML polarization-locked vector second harmonic solitons. (a) The optical spectra before (red line) and after (yellow and green lines) passing through PBS; (b) pulse-trains before (red line) and after (yellow and green lines) passing through PBS (insets: corresponding beam profiles); (c) RF spectrum.

## 3.2 STML GVLVS operation

In our laser, another type of STML vector soliton is also achieved by adjusting the pump power and PCs, as shown in Fig. 5. Similarly, we also first measure the characteristics of this soliton directly output from the laser and before passing through the PBS. The red curve in Fig. 5(a) displays its optical spectrum, featuring a central wavelength of 1560.24 nm and a 3 dB bandwidth of 0.16 nm. The corresponding pulse-train, indicated by the red curve in Fig. 5(b), exhibits a pulse interval of 64.2 ns, indicating that the laser operates in the single-soliton state at the fundamental repetition rate. As shown in Fig. 5(c), the RF spectrum over a span of 500 kHz exhibits a peak with a signal-to-noise ratio of 61.8 dB at the fundamental repetition rate of 15.58 MHz. Moreover, there is no significant intensity modulation in the RF spectrum with a wider span of 1 GHz, as shown in the inset of Fig. 5(c). These indicate that the laser operates stably at the fundamental repetition rate. The measured autocorrelation trace is given in Fig. 5(d), and the pulse width is calculated to be 45.4 ps if the sech$^2$ shape is assumed. The resulting time-bandwidth product is 0.895, indicating that this STML soliton has a certain chirp. The corresponding beam profile in Fig. 5(e) also indicates the presence of high-order transverse modes within this soliton.

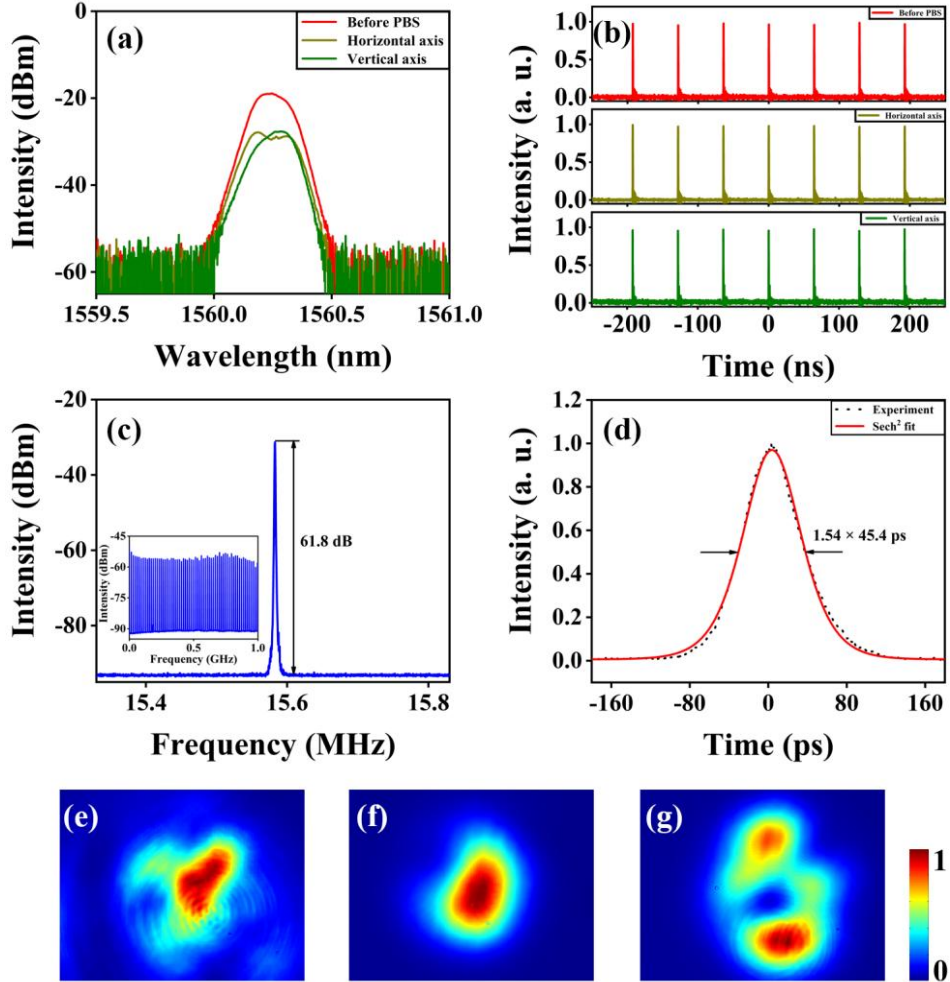

**Figure 5.** Characteristics of the STML group velocity-locked vector single soliton. (a) The optical spectra before (red line) and after (yellow and green lines) passing through extra-cavity PBS; (b) pulse-trains before (red line) and after (yellow and green lines) passing through extra-cavity PBS; (c) RF spectrum with a span of 500 kHz before PBS (inset: RF spectrum with a span of 1 GHz

before PBS); (d) autocorrelation trace before PBS; (e) beam profile before PBS; (f) beam profile of the polarization mode on the horizontal axis; and (g) beam profile of the polarization mode on the vertical axis.

Then, the polarization-resolved measurements of this STML vector soliton are performed at both output ports of the extra-cavity PBS. Its polarization-resolved optical spectra are represented by the yellow and green curves in Fig. 5(a). In addition to the different intensities, the two polarization modes exhibit significantly different spectral profiles. Notably, they also have distinct peak wavelengths and 3 dB spectral bandwidths, both of which differ from those before passing through the PBS. The peak wavelength and 3 dB spectral bandwidth of the polarization mode on the horizontal axis are 1560.19 nm and 0.21 nm, respectively, while those of the polarization mode on the vertical axis are 1560.29 nm and 0.14 nm. The pulse-trains of the two polarization modes are both single-pulse-trains with uniform intensities, as indicated by the yellow and green lines in Fig. 5(b). The corresponding beam profiles are measured and given in Fig. 5(f) and (g). It can be observed that the beam profiles of the two polarization modes are significantly different, with the beam profile of the horizontal polarization mode being closer to the spatial intensity distribution of the fundamental mode than that of the vertical polarization mode. This is to say that different polarization modes possess markedly different transverse mode compositions and relative power ratios, resulting in the modal dispersion between them. Consequently, the formation of this STML vector soliton also necessitates the compensation of modal dispersion between different polarization modes. This STML vector soliton exhibits the same characteristics in the polarization-resolved optical spectra and pulses as the GVLVS in single-mode fiber lasers. However, its two polarization modes also contain distinct transverse mode compositions and relative power ratios, involving locking of multiple transverse modes. Therefore, we refer to this type of STML vector soliton as the STML GVLVS. By drawing an analogy with the GVLVS in a single-mode fiber laser, the difference in the peak wavelengths of the two polarization modes within this STML GVLVS can be attributed to the inability to achieve a delicate balance among the various effects involved in the formation of the STML vector soliton. Therefore, to lock different polarization modes and form the STML vector soliton, additional chromatic dispersion between the two polarization modes needs to be introduced by shifting their peak wavelengths.

Owing to the multimode interference filtering effect and the filtering effect of the Sagnac loop in the cavity, both wavelength-tunable STML PLVSs and STML GVLVSs can be achieved in our laser by adjusting the pump power and PCs. In our laser, both the STML PLVSs and the STML GVLVSs exhibit tunable central wavelengths ranging from approximately 1560 nm to 1567 nm. Although chromatic dispersion, modal dispersion, and polarization modal dispersion vary with the wavelength, adjusting the pump power and PCs also alters nonlinearity, spectral filtering effect, spatial filtering effect, saturable absorption effect, gain, and loss. Therefore, during the complex dynamic self-organization process of STML vector solitons, these effects can reach a new balance, giving rise to STML vector solitons with distinct characteristics. It is noteworthy that as the central wavelength increases, it becomes more difficult to form both types of STML vector solitons, and the stability of the formed solitons deteriorates. This is attributed to the limited operational bandwidth of the SESAM. Its central operational wavelength is 1550 nm, and its full width at half maximum is 15 nm, so it preferentially supports the formation of STML vector solitons near 1550 nm. Nevertheless, the formation of both STML PLVSs and STML GVLVSs at different wavelengths demonstrates that soliton trapping remains a universal phenomenon for vector solitons, even in the more complex STML fiber lasers.

  Furthermore, STML group velocity-locked vector multiple solitons can also be easily achieved in our laser. Keeping the settings of PCs identical to those in Fig. 5, the STML group velocity-locked vector dual solitons, triple solitons, and quadruple solitons are obtained solely by increasing the pump power, as shown in Fig. 6. Figures 6(a) and (b) present the characteristics of the STML group velocity-locked vector dual solitons at a pump power of 210 mW. Before passing through the PBS, the peak wavelength and the 3 dB bandwidth of the optical spectrum are 1560.24 nm and 0.16 nm, as indicated by the red curve in Fig. 6(a). After passing through the PBS, the optical spectra for horizontal and vertical polarization modes exhibit distinct intensities and profiles, as indicated by the yellow and green curves in Fig. 6(a). Additionally, they exhibit different peak wavelengths and 3 dB spectral bandwidths, with the peak wavelengths and 3 dB spectral bandwidths of 1560.20 nm and 0.10 nm for the horizontal polarization mode and 1560.26 nm and 0.16 nm for the vertical polarization mode, respectively. The pulse-trains before and after PBS are shown in Fig. 6(b). There are two pulses with a pulse interval of 21.2 ns within a round-trip time of the cavity, and the pulse intensity remains uniform for each round trip. The corresponding beam profiles shown in the insets of Fig. 6(b) are different, indicating that the formation of the STML group velocity-locked vector dual soliton also involves the compensation of modal dispersion between different polarization modes. When the pump power is increased to 227 mW, STML group velocity-locked vector dual solitons further split into triple solitons, as shown in Figs. 6(c) and (d). Before passing through the PBS, the peak wavelength remains 1560.24 nm, and the 3 dB spectral bandwidth decreases to 0.13 nm, as shown by the red curve in Fig. 6(c). After passing through the PBS, the peak wavelength and 3 dB spectral bandwidth of the horizontal polarization mode are 1560.20 nm and 0.15 nm, while those of the vertical polarization mode are 1560.25 nm and 0.13 nm, respectively, as shown by the yellow and green curves in Fig. 6(c). The pulse-trains before and after PBS are shown in Fig. 6(d). There are three pulses within a round-trip time of the cavity, and the pulse intensity remains uniform for each round trip. The corresponding beam profiles in the insets of Fig. 6(d) also exhibit distinct spatial intensity distributions. Further increasing the pump power to 319 mW, the STML group velocity-locked vector quadruple solitons is obtained, as shown in Figs. 6(e) and (f). Before passing through the PBS, the peak wavelength is blueshifted to 1560.17 nm, while the 3 dB spectral bandwidth remains 0.13 nm, as shown by the red curve in Fig. 6(e). After passing through the PBS, the peak wavelength and 3 dB spectral bandwidth of the horizontal polarization mode are 1560.15 nm and 0.16 nm, while those of the vertical polarization mode are 1560.19 nm and 0.15 nm, respectively, as shown by the yellow and green curves in Fig. 6(e). The pulse-trains before and after PBS are shown in Fig. 6(f). There are four pulses within a round-trip time of the cavity, and the pulse intensity remains uniform for each round trip. The corresponding beam profiles in the insets of Fig. 6(f) are also distinct.

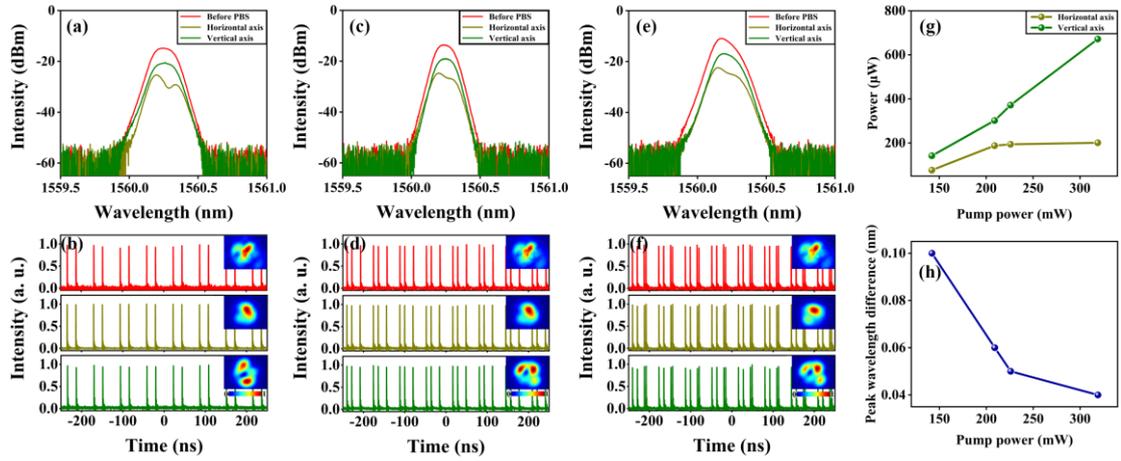

**Figure 6.** Characteristics of the STML group velocity-locked vector multiple solitons obtained solely by adjusting the pump power. (a), (c), and (e) The optical spectra before (red line) and after (yellow and green lines) passing through PBS of dual solitons, triple solitons, and quadruple solitons, respectively; (b), (d), and (f) pulse-trains before (red line) and after (yellow and green lines) passing through PBS of dual solitons, triple solitons, and quadruple solitons, respectively (insets: corresponding beam profiles); (g) output powers of the two polarization modes within multiple solitons versus the pump power; and (h) peak wavelength difference between the two polarization modes within multiple solitons versus the pump power.

By comparing the characteristics of these STML group velocity-locked vector multiple solitons with those of the single soliton in Fig. 5, it can be observed that the increases in spectral intensity are distinct for the two polarization modes with increasing pump power. Thus, we measured the output powers of the two polarization modes within these STML GVLVSs, and their evolutions with pump power are presented in Fig. 6(g). Although the output powers of both polarization modes increase with the pump power, the difference in their output powers grows larger. This suggests that there exists a difference in gain acquisition between the two polarization modes, evidence of gain competition between two polarization modes. Furthermore, the peak wavelengths of the two polarization modes also vary with the increase of pump power. Figure 6(h) presents the evolution of the peak wavelength difference between the two polarization modes with pump power. As the pump power increases, the soliton gradually splits, and the peak wavelength difference between two polarization modes keeps decreasing. This phenomenon may stem from the variations in various effects involved in the formation of STML vector solitons. As the pump power increases, a delicate balance among other effects gradually becomes achievable. Thus, a smaller additional chromatic dispersion needs to be introduced between the two polarization modes by shifting their peak wavelength, resulting in a smaller peak wavelength difference.

In particular, an invisible periodic variation in the beam profile of the STML vector solitons is observed during the soliton splitting process. As observed in Figs. 5(e)-(g) as well as the insets of Figs. 6(b), (d), and (f), during the soliton splitting process induced by the increase in pump power, although the beam profile before passing through the PBS remains almost unchanged, the beam profiles of the two polarization modes vary and alternate. Specifically, when the pump power increases to 210 mW and the single soliton splits into the dual solitons, the beam profile of the horizontal polarization mode varies, while that of the vertical polarization mode remains almost unchanged. When the pump power continues to increase to 227 mW and the dual solitons further split into the triple solitons, the beam profile of the horizontal polarization mode remains almost unchanged, while that of the vertical polarization mode undergoes a significant change. As the pump power increases further to 319 mW and the triple solitons split into the quadruple solitons, the beam profile of the horizontal polarization mode varies, while that of the vertical polarization mode remains almost unchanged. During this soliton splitting process, the beam profile before passing through the PBS always remains almost unchanged, although the beam profiles of two polarization modes vary alternately. Therefore, this variation in the beam profile cannot be observed without polarization-resolved measurements, and this phenomenon is an invisible periodic variation in the beam profile of the STML vector solitons. Since the quintuple solitons are still not realized at relatively higher output power when the pump power was further increased, we do not further increase the pump power to achieve quintuple solitons for the purpose of preventing SESAM damage.

By further adjusting the pump power and PCs, the STML group velocity-locked vector second harmonic solitons are also achieved, as shown in Fig. 7. The optical spectra measured before and after the PBS are presented in Fig. 7(a). Before passing through the PBS, the peak wavelength is 1560.49 nm and the 3 dB spectral bandwidth is 0.18 nm. However, after passing through the PBS, the peak wavelength and 3 dB spectral bandwidth of the horizontal polarization mode are 1560.43 nm and 0.14 nm, while those of the vertical polarization mode are 1560.52 nm and 0.19 nm, respectively. The pulse-trains measured before and after the PBS exhibit a uniform intensity, as shown in Fig. 7(b). Moreover, their pulse intervals are 32.1 ns, indicating that the laser operates in the state of second harmonic solitons. The corresponding beam profiles shown in the insets of Fig. 7(b) exhibit distinct spatial intensity distributions. Interestingly, the horizontal polarization mode exhibits a four-lobe-shaped beam profile resembling the beam profile of the $LP_{21}$ mode to some extent, while the vertical polarization mode demonstrates a beam profile approximating the superposition of the $LP_{01}$ and $LP_{11}$ modes to some extent. Thus, the formation of the STML group velocity-locked vector second harmonic solitons also involves the compensation of modal dispersion between different polarization modes. Figure 7(c) presents the measured RF spectrum. There is a relatively weak peak at the fundamental repetition rate of 15.58 MHz, while a peak with 40.4 dB stronger intensity is located at the repetition rate of 31.16 MHz, further confirming the state of the second harmonic solitons. Similarly, by carefully adjusting the pump power and PCs, the STML group velocity-locked vector third-harmonic solitons and fourth-harmonic solitons are realized in our laser.

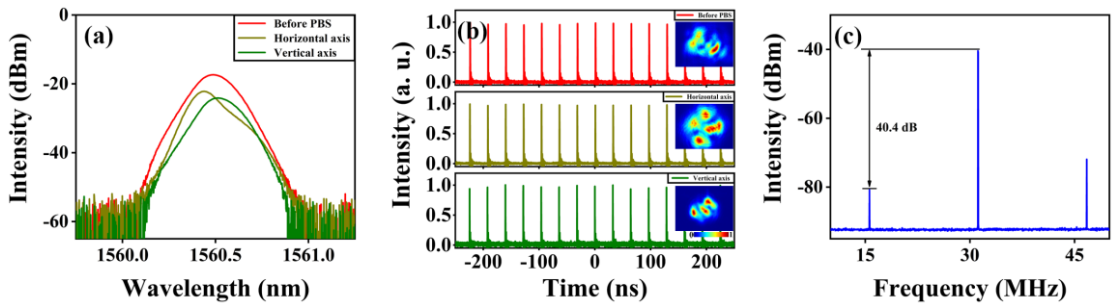

**Figure 7.** Characteristics of the STML group velocity-locked vector second harmonic solitons. (a) The optical spectra before (red line) and after (yellow and green lines) passing through PBS; (b) pulse-trains before (red line) and after (yellow and green lines) passing through PBS (insets: corresponding beam profiles); (c) RF spectrum.

## 4. Discussion

In our laser, two novel types of vector solitons associated with transverse modes, namely the STML PLVS and the STML GVLVS, are achieved. Both in the STML PLVS and the STML GVLVS, their two polarization modes possess distinct transverse mode compositions and relative power ratios, resulting in the modal dispersion between them. In order to realize the locking of the two polarization modes, this modal dispersion needs to be counteracted. Therefore, the formation of these two types of STML vector solitons also involves the compensation of modal dispersion between different polarization modes. Based on the formation mechanisms of vector solitons in single-mode fiber lasers and STML scalar solitons in multimode fiber lasers, the formation mechanism of STML vector solitons can be summarized as a delicate balance among chromatic dispersion, modal dispersion, polarization modal dispersion, nonlinearity, spectral filtering effect, spatial filtering effect, saturable absorption effect, gain, and loss. In STML polarization-locked vector single soliton, multiple solitons, and harmonic solitons, the two polarization modes share identical peak wavelength. However, in STML group velocity-locked vector single soliton, multiple solitons, and harmonic solitons, the peak wavelengths of the two polarization modes are different. Moreover, the peak wavelength difference between the two polarization modes varies in different STML group velocity-locked vector solitons. The difference between these two types of STML vector solitons in terms of the peak wavelength disparity between the two polarization modes can be attributed to the variations in the various effects involved in the formation of STML vector solitons. After achieving the steady STML vector soliton, adjusting the pump power and PCs alters these effects, disrupting the original balance and necessitating the establishment of a new balance for vector soliton formation. During this complex dynamic soliton self-organization process, the peak wavelengths of the two polarization modes can either be identical or different to counteract the variations in these effects. When these effects can still achieve a delicate balance, there is no need to introduce additional chromatic dispersion between the two polarization modes by shifting their peak wavelengths to form the vector soliton. This results in the two polarization modes having the same peak wavelength, forming an STML PLVS. Conversely, if these effects fail to achieve a delicate balance, additional chromatic dispersion between the two polarization modes must be introduced by shifting their peak wavelengths. This results in distinct peak wavelengths between the two polarization modes, and an STML GVLVS is formed. These characteristics demonstrate that soliton trapping remains a universal phenomenon for vector solitons, even in the more intricate STML fiber lasers.

Notably, in the STML GVLVS, some interesting polarization-dependent phenomena are observed during the soliton splitting induced by increasing the pump power while keeping the PCs unchanged. During this process, as the pump power increases, the output powers of both polarization modes increase. However, their difference grows larger, suggesting that there exists a difference in gain acquisition between the two polarization modes, evidence of gain competition between two polarization modes. Moreover, the peak wavelength difference between two polarization modes decreases as the number of solitons increases, which may be due to the variations in the various effects involved in the formation of STML vector solitons when the pump power is increased. In particular, we also observe an invisible periodic variation in the beam profile of the STML vector solitons during the soliton splitting process. As the single soliton gradually splits into quadruple solitons, the beam profile of the horizontal polarization mode first varies, then remains almost unchanged, and finally varies again. Conversely, the beam profile of the vertical polarization mode first remains almost unchanged, then undergoes a significant change, and finally remains almost unchanged again. However, the beam profile before passing through the PBS always remains almost unchanged during this process. In order to avoid damaging the SESAM, the pump power was not further increased to obtain vector solitons with more solitons for an in-depth investigation. The further study could be conducted by employing the saturable absorbers with higher damage thresholds, such as nonlinear polarization rotation technique [5] and nonlinear amplifying loop mirror technique [28].

Benefiting from the multimode interference filtering effect induced by the difference in fiber core diameters and the filtering effect of the Sagnac loop in the cavity, the wavelengths of both STML PLVSs and STML GVLVSs can be tuned from approximately 1560 nm to 1567 nm. Since the SESAM used in our laser has a central operational wavelength of 1550 nm and a limited operational bandwidth, it becomes more challenging to achieve both types of STML vector solitons at a longer central wavelength, and the stability of the formed solitons deteriorates. To achieve STML vector solitons operating at other wavelengths and exhibiting a larger wavelength tunable range, SESAMs or other saturable absorbers with different central operational wavelengths and larger operational bandwidth can be employed. Although distinct beam profiles for the two polarization modes are observed in both types of STML vector solitons, we are unable to obtain detailed information about the transverse modes contained in the two polarization modes due to the limitations of the current measurement equipment and technology. Although transverse mode compositions and relative power ratios within STML solitons can be obtained by employing the delay-scanning off-axis digital holography approach, it is rather complicated and costly [29]. Very recently, the spatiotemporal dispersive Fourier transform technique has been proposed to measure the transverse-mode-resolved spectrum, transverse mode compositions, and relative power ratios within STML solitons [30]. However, it requires a very long fiber identical to the output fiber of the laser, and the output fiber of our laser is expensive, limiting the use of the technique. Therefore, it is necessary to develop new mode-resolved measurement techniques to further study and reveal the detailed transverse mode characteristics of the two polarization modes in STML vector solitons.

## 5. Conclusion

To summarize, we investigate the vector characteristics of the STML fiber laser and demonstrate two novel types of vector solitons associated with transverse modes. In both types of STML vector solitons, distinct transverse mode compositions and relative power ratios for different polarization modes are observed. However, different polarization modes share identical peak wavelengths in STML PLVSs, while they have different peak wavelengths in STML GVLVSs. By optimization of the pump power and PCs, both STML PLVS and STML GVLVS can operate in the single soliton, multiple solitons, and harmonic solitons. During the soliton splitting process of the STML GVLVSs, gain competition between polarization modes is observed. In addition, as the number of solitons increases, the peak wavelength difference between the two polarization modes decreases. Particularly, during this process, the beam profiles of the two polarization modes vary and alternate, although the beam profile before passing through the PBS remains almost unchanged. Moreover, due to the multimode interference filtering effect and the filtering effect of the Sagnac loop in the cavity, the wavelengths of both STML PLVSs and STML GVLVSs can be tuned from approximately 1560 nm to 1567 nm. The formation of STML vector solitons demonstrates that soliton trapping remains a universal phenomenon for vector solitons even in the more intricate STML fiber lasers, and the obtained results reveal the vector characteristics of STML fiber lasers, contributing to a deep understanding of their nonlinear phenomena.

**Acknowledgement**

This work was supported by the National Natural Science Foundation of China (Grant Nos. 62375091, 92050101); The Natural Science Foundation of Guangdong Province (2021A1515011608, 2023A1515011870).